\documentclass[fleqn]{llncs} \usepackage{llncs_ext}

\usepackage{amsmath,amssymb,atmath}               

\newcommand{\Half}[1][1]{\frac{#1}{2}}
\newcommand{\tHalf}[1][1]{\tfrac{#1}{2}}

\usepackage[pdftex]{graphicx}
\usepackage{verbatim,url}

\begin{document}

\begin{frontmatter}

\title{Faster subsequence recognition\\ in compressed strings%
}
\myself\maketitle

\begin{abstract}

Computation on compressed strings is one of the key approaches 
to processing massive data sets.
We consider local subsequence recognition problems
on strings compressed by straight-line programs (SLP),
which is closely related to Lempel--Ziv compression.
For an SLP-compressed text of length $\bar m$,
and an uncompressed pattern of length $n$,
C{\'e}gielski et al.\ gave an algorithm for local subsequence recognition 
running in time $O(\bar mn^2 \log n)$.
We improve the running time to $O(\bar mn^{1.5})$.
Our algorithm can also be used to compute the longest common subsequence
between a compressed text and an uncompressed pattern 
in time $O(\bar mn^{1.5})$;
the same problem with a compressed pattern is known to be NP-hard.

\end{abstract}

\end{frontmatter}

\section{Introduction}
\label{s-intro}

Computation on compressed strings is one of the key approaches 
to processing massive data sets.
It has long been known that certain algorithmic problems  
can be solved directly on a compressed string,
without first decompressing it;
see \cite{Cegielski+:06,Lifshits_Lohrey:06} for references.

One of the most general string compression methods
is compression by straight-line programs (SLP) \cite{Rytter:03}.
In particular, SLP compression captures the well-known LZ and LZW algorithms
\cite{Ziv_Lempel:77,Ziv_Lempel:78,Welch:84}.
Various pattern matching problems on SLP-compressed strings 
have been studied; see e.g.\ \cite{Cegielski+:06} for references.
C{\'e}gielski et al.\ \cite{Cegielski+:06} considered
subsequence recognition problems on SLP-compressed strings.
For an SLP-compressed text of length $\bar m$,
and an uncompressed pattern of length $n$,
they gave several algorithms for global and local subsequence recognition,
running in time $O(\bar mn^2 \log n)$.

In this paper, we improve
on the results of \cite{Cegielski+:06} as follows.
First, we describe a simple folklore algorithm 
for global subsequence recognition
on an SLP-compressed text, running in time $O(\bar mn)$.
Then, we consider the more general 
partial semi-local longest common subsequence (LCS) problem,
which consists in computing implicitly
the LCS between the compressed text 
and every substring of the uncompressed pattern.
The same problem with a compressed pattern is known to be NP-hard.
For the partial semi-local LCS problem,
we propose a new algorithm, running in time $O(\bar mn^{1.5})$.
Our algorithm is based on 
the partial highest-score matrix multiplication technique
presented in \cite{Tiskin:llcs-survey}.
We then extend this method to the several versions 
of local subsequence recognition considered in \cite{Cegielski+:06},
for each obtaining an algorithm 
running in the same asymptotic time $O(\bar mn^{1.5})$.

This paper is a sequel to papers \cite{Tiskin:JDA_CSR,Tiskin:llcs-survey};
we recall most of their relevant material here for completeness.

\section{Subsequences in compressed text}
\label{c-slp}

We consider strings of characters from a fixed finite alphabet,
denoting string concatenation by juxtaposition.
Given a string, we distinguish between its contiguous \emph{substrings},
and not necessarily contiguous \emph{subsequences}.
Special cases of a substring 
are \emph{a prefix} and \emph{a suffix} of a string.
Given a string $a$ of length $m$,
we use the \emph{take/drop notation} of \cite{Watson_Zwaan:96},
$a \ltake k$, $a \ldrop k$, $a \rtake k$, $a \rdrop k$,
to denote respectively its 
prefix of length $k$, suffix of length $m-k$,
suffix of length $k$, and prefix of length $m-k$.
For two strings $a= \alpha_1 \alpha_2 \ldots \alpha_m$ 
and $b= \beta_1 \beta_2 \ldots \beta_n$
of lengths $m$, $n$ respectively,
the \emph{longest common subsequence (LCS) problem}
consists in computing the length of the longest string
that is a subsequence of both $a$ and $b$.
We will call this length the \emph{LCS score} of the strings.

Let $T$ be a string of length $m$ (typically large).
String $T$ will be represented implicitly 
by a \emph{straight-line program (SLP)} of \emph{length $\bar m$},
which is a sequence of $\bar m$ \emph{statements}.
Each statement $r$, $1 \leq r \leq \bar m$,
has either the form $T_r = \alpha$, where $\alpha$ is an alphabet character,
or the form $T_r = T_s T_t$, where $1 \leq s,t < r$.
We identify every symbol $T_r$ with the string it represents;
in particular, we have $T=T_{\bar m}$.
Note that $m \geq \bar m$, and that in general the uncompressed text length $m$ 
can be exponential in the SLP-compressed length $\bar m$.

Our goal is to design efficient algorithms on SLP-compressed texts.
While we do not allow text decompression
(since, in the worst case, this could be extremely inefficient),
we will assume that standard arithmetic operations on integers up to $m$
can be performed in constant time.
This assumption is necessary, since the counting version of our problem
produces a numerical output that may be as high as $O(m)$.
The same assumption on the computation model
is made implicitly in \cite{Cegielski+:06}.

The LCS problem on uncompressed strings is a classical problem;
see e.g.\ \cite{Crochemore_Rytter:94,Gusfield:97}
for the background and references.
Given input strings of lengths $m$, $n$, the LCS problem
can be solved in time $O\bigpa{\frac{mn}{log(m+n)}}$,
assuming $m$ and $n$ are reasonably close
\cite{Masek_Paterson:80,Crochemore+:03_SIAM}.
The LCS problem on two SLP-compressed strings
is considered in \cite{Lifshits_Lohrey:06},
and proven to be NP-hard.
In this paper, we consider the LCS problem on two input strings,
one of which is SLP-compressed and the other uncompressed.
This problem can be regarded as a special case 
of computing the edit distance between
a context-free language given by a grammar of size $\bar m$,
and a string of size $n$.
For this more general problem, Myers \cite{Myers:95}
gives an algorithm running in time 
$O(\bar m n^3 + \bar m \log \bar m \cdot n^2)$.

From now on, we will assume that string $T$ (the \emph{text string})
of length $m$ is represented by an SLP of length $\bar m$,
and that string $P$ (the \emph{pattern string}) of length $n$
is represented explicitly.
Following \cite{Cegielski+:06,Lifshits_Lohrey:06},
we study the problem of recognising in $T$ subsequences identical to $P$,
which is closely related to the LCS problem.

\begin{mydefinition}
\label{def-global}
The \emph{(global) subsequence recognition problem}
consists in deciding whether string $T$ 
contains string $P$ as a subsequence.
\end{mydefinition}

The subsequence recognition problem on uncompressed strings 
is a classical problem, considered e.g.\ in \cite{Aho+:76}
as the ``subsequence matching problem''.
The subsequence recognition problem on an SLP-compressed text
is considered in \cite{Cegielski+:06} as Problem 1,
with an algorithm running in time $O(\bar m n^2 \log n)$.

In addition to global subsequence recognition,
it is useful to consider text subsequences locally,
i.e.\ in substrings of $T$.
In this context, we will call the substrings of $T$ \emph{windows}.
We will say that string $a$ contains string $b$ 
\emph{minimally} as a subsequence,
if $b$ is a subsequence in $a$, but not in any proper substring of $a$.
Even with this restriction on subsequence containment,
the number of substrings in $T$ containing $P$ 
minimally as a subsequence may be as high as $O(m)$,
so just listing them all may require time exponential in $\bar m$.
The same is true if, instead of minimal substrings,
we consider all substrings of $T$ of a fixed length.
Therefore, it is sensible to define
local subsequence recognition as a family of counting problems
\begin{mydefinition}
\label{def-local-minimal}
The \emph{minimal-window subsequence recognition problem} consists
in counting the number of windows in string $T$, 
containing string $P$ minimally as a subsequence.
\end{mydefinition}
\begin{mydefinition}
\label{def-local-fixed}
The \emph{fixed-window subsequence recognition problem} consists
in counting the number of windows of a given length $w$ in string $T$, 
containing string $P$ as a subsequence.
\end{mydefinition}

The minimal-window and fixed-window subsequence recognition problems
on uncompressed strings
are considered in \cite{Das+:97} as ``episode matching problems''
(see also \cite{Cegielski+:06_IPL} and references therein).
The same problems on an SLP-compressed text and an uncompressed pattern
are considered in \cite{Cegielski+:06} 
as Problems 2, 3 (a special case of 2) and 4.
Additionally, the same paper considers
the \emph{bounded minimal-window subsequence recognition problem}
(counting the number of windows in $T$ of length at most $w$ 
containing $P$ minimally as a subsequence) as Problem 5.
For all these problems, paper \cite{Cegielski+:06} 
gives algorithms running in time $O(\bar mn^2 \log n)$.

\section{Semi-local longest common subsequences}
\label{s-llcs}

In this section and the next, we recall the algorithmic framework 
developed in \cite{Tiskin:JDA_CSR,Tiskin:llcs-survey}.
This framework is subsequently used to solve 
the compressed subsequence recognition problems
introduced in the previous section.

\newcommand{\tl}{\vartriangleleft}
\newcommand{\tleq}{\trianglelefteq}

In \cite{Tiskin:JDA_CSR},
we introduced the following problem.
\begin{mydefinition}
\label{def-semilocal}
The \emph{all semi-local LCS problem} consists in computing the LCS scores
on substrings of strings $a$ and $b$ as follows:
\begin{itemize}
\item the \emph{all string-substring LCS problem}:
$a$ against every substring of $b$;
\item the \emph{all prefix-suffix LCS problem}:
every prefix of $a$ against every suffix of $b$;
\item symmetrically, the \emph{all substring-string LCS problem}
and the \emph{all suffix-prefix LCS problem},
defined as above but with the roles of $a$ and $b$ exchanged.
\end{itemize}
\end{mydefinition}
It turns out that this is a very natural and useful 
generalisation of the LCS problem.
%

In addition to standard integer indices 
$\ldots, -2, -1, 0, 1, 2, \ldots$,
we use \emph{odd half-integer} indices 
$\ldots, -\Half[5], -\Half[3], -\Half[1], 
 \Half[1], \Half[3], \Half[5], \ldots$.
For two numbers $i$, $j$, we write $i \tleq j$ if $j-i \in \{0,1\}$,
and $i \tl j$ if $j-i = 1$.
We denote
\begin{gather*}
\bra{i:j} = \{i, i+1, \ldots, j-1, j\}\qquad 
\ang{i:j} = \bigbrc{i+\tHalf[1], i+\tHalf[3], \ldots, 
                    j-\tHalf[3], j-\tHalf[1]}
\end{gather*}
To denote infinite intervals of integers and odd half-integers,
we will use $-\infty$ for $i$ and $+\infty$ for $j$ where appropriate.
For both interval types $\bra{i:j}$ and $\ang{i:j}$,
we call the difference $j-i$ interval \emph{length}.

We will make extensive use of finite and infinite matrices,
with integer elements and integer or odd half-integer indices.
A \emph{permutation matrix} is a (0,1)-matrix 
containing exactly one nonzero in every row and every column.
An \emph{identity matrix} is a permutation matrix $I$,
such that $I(i,j)=1$ if $i=j$, and $I(i,j)=0$ otherwise.
Each of these definitions applies to both finite and infinite matrices.

From now on, instead of ``index pairs corresponding to nonzeros'',
we will write simply ``nonzeros'', where this does not lead to confusion.
A finite permutation matrix can be represented by its nonzeros.
When we deal with an infinite matrix,
it will typically have a finite non-trivial core,
and will be trivial (e.g.\ equal to an infinite identity matrix)
outside of this core.
An infinite permutation matrix with finite non-trivial core
can be represented by its core nonzeros.


Let $D$ be an arbitrary numerical matrix
with indices ranging over $\ang{0:n}$.
Its \emph{distribution matrix}, with indices ranging over $\bra{0:n}$,
is defined by
%
\begin{gather*}
\label{eq-distribution}
d(i_0,j_0) = \sum D(i,j) \qquad i \in \ang{i_0:n}, j \in \ang{0:j_0}
\end{gather*}
for all $i_0,j_0 \in \bra{0:n}$.
We have
%
\begin{gather*}
D(i,j) = 
d(i-\tHalf,j+\tHalf) - d(i-\tHalf,j-\tHalf) -
d(i+\tHalf,j+\tHalf) + d(i+\tHalf,j-\tHalf)
\end{gather*}

When matrix $d$ is a distribution matrix of $D$,
matrix $D$ is called the \emph{density matrix} of $d$.
The definitions of distribution and density matrices 
extend naturally to infinite matrices.
We will only deal with distribution matrices 
where all elements are defined and finite.

We will use the term \emph{permutation-distribution matrix}
as an abbreviation of ``distribution matrix of a permutation matrix''.

\section{Algorithmic techniques}
\label{c-techniques}

The rest of this paper is based on the framework 
for the all semi-local LCS problem 
developed in \cite{Tiskin:JDA_CSR,Tiskin:llcs-survey}.
For completeness, we recall most background definitions and results 
from \cite{Tiskin:JDA_CSR}, omitting the proofs.

\subsection{Dominance counting}
\label{s-dominance}

It is well-known that an instance of the LCS problem 
can be represented by a dag (directed acyclic graph)
on an $m \times n$ grid of nodes,
where character matches correspond to edges scoring 1,
and mismatches to edges scoring 0.

\begin{mydefinition}
Let $m,n \in \Nat$.
An \emph{alignment dag} $G$ is a weighted dag, 
defined on the set of nodes $v_{l,i}$, 
$l \in \bra{0:m}$, $i \in \bra{0:n}$.
The edge and path weights are called \emph{scores}.
For all $l \in \bra{1:m}$, $i \in \bra{1:n}$:
\begin{itemize}
\item horizontal edge $v_{l,i-1} \to v_{l,i}$
and vertical edge $v_{l-1,i} \to v_{l,i}$ 
are both always present in $G$ and have score $0$;
\item diagonal edge $v_{l-1,i-1} \to v_{l,i}$ 
may or may not be present in $G$; if present, it has score $1$.
\end{itemize}
Given an instance of the all semi-local LCS problem,
its \emph{corresponding alignment dag} is an $m \times n$ alignment dag,
where the diagonal edge $v_{l-1,i-1} \to v_{l,i}$ is present,
iff $\alpha_i = \beta_j$.
\end{mydefinition}

\begin{figure}[tb]
\centering
\includegraphics{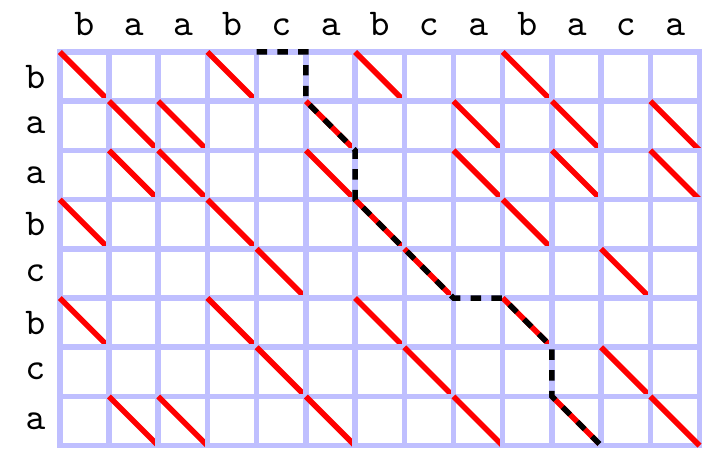}
\caption{\label{f-align} An alignment dag and a highest-scoring path}
\end{figure}
Figure~\ref{f-align} shows the alignment dag corresponding to strings 
$a = \text{``\texttt{baabcbca}''}$, $b = \text{``\texttt{baabcabcabaca}''}$
(an example borrowed from \cite{Alves+:05}).

Common string-substring, suffix-prefix, prefix-suffix, and substring-string
subsequences correspond, respectively, to paths of the following form
in the alignment dag:
\begin{gather}
v_{0,i} \pathto v_{m,i'},\quad
v_{l,0} \pathto v_{m,i'},\quad
v_{0,i} \pathto v_{l',n},\quad
v_{l,0} \pathto v_{l',n},
\label{eq-paths}
\end{gather}
where $l,l' \in [0:m]$, $i,i' \in [0:n]$.
The length of each subsequence is equal 
to the score of its corresponding path.

The solution to the all semi-local LCS problem is equivalent 
to finding the score of a highest-scoring path
of each of the four types \eqref{eq-paths}
between every possible pair of endpoints.

To describe our algorithms, 
we need to modify the definition of the alignment dag
by embedding the finite grid of nodes into in an infinite grid.

\begin{mydefinition}
\label{def-align-ext}
Given an $m \times n$ alignment dag $G$,
its \emph{extension} $G^+$ is an infinite weighted dag, 
defined on the set of nodes $v_{l,i}$, $l,i \in \bra{-\infty:+\infty}$ 
and containing $G$ as a subgraph.
For all $l,i \in \bra{-\infty:+\infty}$:
\begin{itemize}
\item horizontal edge $v_{l,i-1} \to v_{l,i}$ 
and vertical edge $v_{l-1,i} \to v_{l,i}$ 
are both always present in $G^+$ and have score $0$;
\item when $l \in [1:m]$, $i \in [1:n]$,
diagonal edge $v_{l-1,i-1} \to v_{l,i}$
is present in $G^+$ iff it is present in $G$; if present, it has score $1$;
\item otherwise, diagonal edge $v_{l-1,i-1} \to v_{l,i}$
is always present in $G^+$ and has score $1$.
\end{itemize}
An infinite dag that is an extension of some (finite) alignment dag
will be called an \emph{extended alignment dag}.
When dag $G^+$ is the extension of dag $G$, 
we will say that $G$ is the \emph{core} of $G^+$.
Relative to $G^+$, we will call the nodes of $G$ \emph{core nodes}.
\end{mydefinition}

By using the extended alignment dag representation,
the four path types \eqref{eq-paths} can be reduced to a single type,
corresponding to the all string-substring 
(or, symmetrically, substring-string) LCS problem
on an extended set of indices.

\begin{mydefinition}
\label{def-matrix}
Given an $m \times n$ alignment dag $G$,
its \emph{extended highest-score matrix} 
is an infinite matrix defined by
\begin{align}{}
&A(i,j) = \max \mathit{score}(v_{0,i} \pathto v_{m,j}) &
  &i,j \in \bra{-\infty:+\infty} 
\label{eq-Ss}
\end{align}
where the maximum is taken across all paths 
between the given endpoints in the extension $G^+$.
If $i=j$, we have $A(i,j)=0$. 
By convention, if $j < i$, then we let $A(i,j)=j-i<0$.
\end{mydefinition}
In Figure~\ref{f-align}, the highlighted path has score $5$, 
and corresponds to the value $A(4,11)=5$,
which is equal to the LCS score of string $a$ 
and substring $b' = \text{``\texttt{cabcaba}''}$.

In this paper, we will deal almost exclusively with extended
(i.e.\ finitely represented, but conceptually infinite)
alignment dags and highest-score matrices.
From now on, we omit the term ``extended'' for brevity,
always assuming it by default.

The maximum path scores for each of the four path types \eqref{eq-paths}
can be obtained from the highest-score matrix \eqref{eq-Ss}
as follows:
\begin{gather*}
\max \mathit{score}(v_{0,j} \pathto v_{m,j'}) = A(j,j') \\
\max \mathit{score}(v_{i,0} \pathto v_{m,j'}) = A(-i,j')-i \\
\max \mathit{score}(v_{0,j} \pathto v_{i',n}) = A(j,m+n-i')-m+i' \\
\max \mathit{score}(v_{i,0} \pathto v_{i',n}) = A(-i,m+n-i')-m-i+i'
\end{gather*}
where $i,i' \in [0:m]$, $j,j' \in [0:n]$,
and the maximum is taken across all paths between the given endpoints.

\begin{mydefinition}
\label{def-critical}
An odd half-integer point $(i,j) \in \ang{-\infty:+\infty}^2$ 
is called \emph{$A$-critical}, if
\begin{equation*}
A\bigpa{i+\tHalf,j-\tHalf} \tl A\bigpa{i-\tHalf,j-\tHalf} = 
A\bigpa{i+\tHalf,j+\tHalf} =   A\bigpa{i-\tHalf,j+\tHalf}
\end{equation*}
\end{mydefinition}
In particular, 
point $(i,j)$ is never $A$-critical for $i>j$.
When $i=j$, 
point $(i,j)$ is $A$-critical iff $A\bigpa{i-\Half,j+\Half} = 0$.

\begin{mycorollary}
\label{cr-critical}
Let $i,j \in \ang{-\infty:+\infty}$.
For each $i$ (respectively, $j$),
there exists exactly one $j$ (respectively, $i$)
such that the point $(i,j)$ is $A$-critical.
\end{mycorollary}
%
%
\begin{figure}[tb]
\centering
\includegraphics{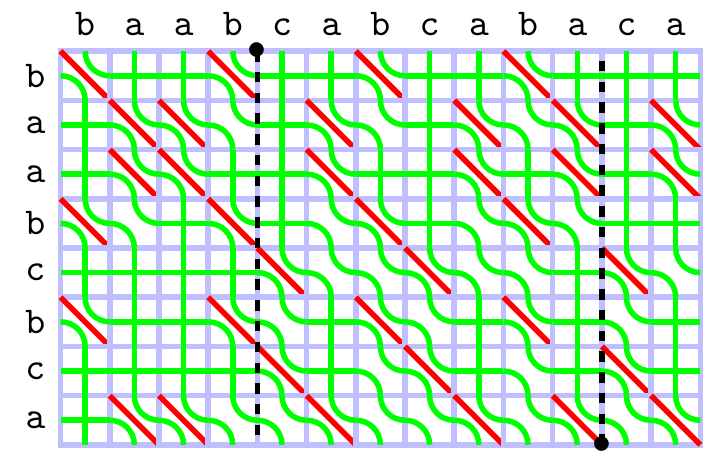}
\caption{\label{f-align-crit} 
An alignment dag and the seaweeds}
\end{figure}
Figure~\ref{f-align-crit} shows the alignment dag of Figure~\ref{f-align}
along with the critical points.
In particular, every critical point $(i,j)$, 
where $i,j \in \ang{0:n}$,
is represented by a \emph{seaweed}%
\footnote{This imaginative term was suggested by Yu.~V.~Matiyasevich.},
originating between the nodes $v_{0,i-\Half}$ and $v_{0,i+\Half}$,
and terminating between the nodes $v_{m,j-\Half}$ and $v_{m,j+\Half}$.
The remaining seaweeds, 
originating or terminating at the sides of the dag,
correspond to critical points $(i,j)$,
where either $i \in \ang{-m:0}$ or $j \in \ang{n:n+m}$ (or both).
In particular, every critical point $(i,j)$, 
where $i \in \ang{-m:0}$ (respectively, $j \in \ang{n:m+n}$)
is represented by a seaweed 
originating between the nodes $v_{-i-\Half,0}$ and $v_{-i+\Half,0}$
(respectively, terminating between the nodes 
$v_{m+n-j-\Half,n}$ and $v_{m+n-j+\Half,n}$).

It is convenient to consider 
the set of $A$-critical points as an infinite permutation matrix.
For all $i,j \in \ang{-\infty:+\infty}$, we define
\begin{gather*}
D_A(i,j)=
\begin{cases}
1 & \text{if $(i,j)$ is $A$-critical}\\
0 & \text{otherwise}
\end{cases}
\end{gather*}
We denote the infinite distribution matrix of $D_A$ by $d_A$,
and consider the following simple geometric relation.
\begin{mydefinition}
\label{def-dominance}
Point $(i_0,j_0)$ \emph{dominates%
\footnote{The standard definition of dominance requires 
$i < i_0$ instead of $i_0 < i$.
Our definition is more convenient in the context of the LCS problem.}} 
point $(i,j)$, if $i_0 < i$ and $j < j_0$.
\end{mydefinition}
Informally, the dominated point is ``below and to the left''
of the dominating point in the highest-score matrix\footnote{%
Note that these concepts of ``below'' and ``left'' 
are relative to the highest-score matrix, and have no connection 
to the ``vertical'' and ``horizontal'' directions in the alignment dag.}.
Clearly, for an arbitrary integer point $(i_0,j_0) \in \bra{-\infty:+\infty}^2$,
the value $d_A(i_0,j_0)$ is the number of (odd half-integer) 
$A$-critical points it dominates.

The following theorem shows that the set of critical points 
defines uniquely a highest-score matrix,
and gives a simple formula for recovering the matrix elements.
\begin{mytheorem}
\label{th-recover}
For all $i_0,j_0 \in \bra{-\infty:+\infty}$, we have 
\begin{gather*}
A(i_0,j_0) = j_0 - i_0 - d_A(i_0,j_0)
\end{gather*} 
\end{mytheorem}
In Figure~\ref{f-align-crit}, critical points dominated by point $(4,11)$
are represented by seaweeds whose both endpoints 
(and therefore the whole seaweed) fit between the two vertical lines,
corresponding to index values $i=4$ and $j=11$.
Note that there are exactly two such seaweeds,
and that $A(4,11)=11-4-2=5$.


By \thref{th-recover}, a highest-score matrix $A$
is represented uniquely by an infinite permutation matrix $D_A$
with odd half-integer row and column indices.
We will call matrix $D_A$ the \emph{implicit representation} of $A$.
From now on, we will refer to the critical points of $A$
as nonzeros (i.e., ones) in its implicit representation.

Recall that outside the core,
the structure of an alignment graph is trivial:
all possible diagonal edges are present in the off-core subgraph.
This property carries over to the corresponding permutation matrix.
\begin{mydefinition}
Given an infinite permutation matrix $D$,
its \emph{core} is a square (possibly semi-infinite) submatrix 
defined by the index range $\bra{i_0:j_0} \times \bra{i_1:j_1}$,
where $j_0-i_0=j_1-i_1$ (as long as both these values are defined),
and such that for all off-core elements $D(i,j)$,
we have $D(i,j)=1$ iff $j-i=j_0-i_0$ and $j-i=j_1-i_1$ 
(in each case, as long as the right-hand side is defined).
\end{mydefinition}
Informally, the off-core part of matrix $D$ 
has nonzeros on the off-core extension of the core's main diagonal.

The following statements are an immediate consequence of the definitions.
\begin{mycorollary}
A core of an infinite permutation matrix
is a (possibly semi-infinite) permutation matrix.
\end{mycorollary}
\begin{mycorollary}
Given an alignment dag $A$ as described above,
the corresponding permutation matrix $D_A$ has core of size $m+n$, 
defined by $i \in \ang{-m,n}$, $j \in \ang{0,m+n}$.
\end{mycorollary}
In Figure~\ref{f-align-crit}, the set of critical points
represented by the seaweeds corresponds precisely
to the set of all core nonzeros in $D_A$.
Note that there are $m+n=8+13=21$ seaweeds in total.

Since only core nonzeros need to be represented explicitly,
the implicit representation of a highest-score matrix
can be stored as a permutation of size $m+n$.
From now on, we will assume this as the default representation
of such matrices.

By \thref{th-recover}, the value $A(i_0,j_0)$
is determined by the number of nonzeros in $D_A$
dominated by $(i_0,j_0)$.
Therefore, an individual element of $A$ can be obtained explicitly
by scanning the implicit representation of $A$ in time $O(m+n)$,
counting the dominated nonzeros.
However, existing methods of computational geometry
allow us to perform this \emph{dominance counting} procedure
much more efficiently, as long as preprocessing 
of the implicit representation is allowed.


\begin{mytheorem}
\label{th-range-tree}
Given the implicit representation $D_A$ of a highest-score matrix $A$,
there exists a data structure which
\begin{itemize}
\item has size $O\bigpa{(m+n) \log (m+n)}$;
\item can be built in time $O\bigpa{(m+n) \log (m+n)}$;
\item allows to query an individual element of $A$
      in time $O\bigpa{\log^2 (m+n)}$.
\end{itemize}
\end{mytheorem}

\subsection{Highest-score matrix multiplication}
\label{s-score-mmult}

A common pattern in many problems on strings
is partitioning the alignment dag into alignment subdags.
Without loss of generality, consider a partitioning 
of an $(M+m) \times n$ alignment dag $G$ 
into an $M \times n$ alignment dag $G_1$ 
and an $m \times n$ alignment dag $G_2$,
where $M \geq m$.
The dags $G_1$, $G_2$ share a horizontal row of $n$ nodes,
which is simultaneously the bottom row of $G_1$ and the top row of $G_2$;
the dags also share the corresponding $n-1$ horizontal edges.
We will say that dag $G$ is the \emph{concatenation}
of dags $G_1$ and $G_2$.
Let $A$, $B$, $C$ denote the highest-score matrices
defined respectively by dags $G_1$, $G_2$, $G$.
Our goal is, given matrices $A$, $B$, to compute matrix $C$ efficiently.
We call this procedure \emph{highest-score matrix multiplication}.

%
%
%
%
\begin{mydefinition}
\label{def-minplus}
Let $n \in \Nat$. 
Let $A$, $B$, $C$ be arbitrary numerical matrices 
with indices ranging over $\bra{0:n}$.
The \emph{$(\min,+)$-product} $A \odot B = C$ is defined by
\begin{align*}
&C(i,k) = \min_j \bigpa{A(i,j) + B(j,k)} &
&i,j,k \in \bra{0:n}
\end{align*}

\end{mydefinition}
\begin{mylemma}[\cite{Tiskin:JDA_CSR}]
\label{lm-m-mmult}
Let $D_A$, $D_B$, $D_C$ be permutation matrices
with indices ranging over $\ang{0:n}$,
and let $d_A$, $d_B$, $d_C$ be their respective distribution matrices.
Let $d_A \odot d_B = d_C$.
Given the nonzeros of $D_A$, $D_B$, the nonzeros of $D_C$
can be computed in time $O\bigpa{n^{1.5}}$ and memory $O(n)$.
\end{mylemma}
\begin{mylemma}[\cite{Tiskin:JDA_CSR}]
\label{lm-m-mmult-inf1}
Let $D_A$, $D_B$, $D_C$ be permutation matrices
with indices ranging over $\ang{-\infty:+\infty}$.
Let $D_A$ (respectively, $D_B$) have semi-infinite core
$\ang{0:+\infty}^2$ (respectively, $\ang{-\infty:n}^2$).
Let $d_A$, $d_B$, $d_C$ be the respective distribution matrices,
and assume $d_A \odot d_B = d_C$.
We have
\begin{align}
\label{lm-m-mmult-inf1-triv1}
&D_A(i,j)=D_C(i,j) & 
&\text{for $i \in \ang{-\infty:+\infty}$, $j \in \ang{n:+\infty}$}\\
\label{lm-m-mmult-inf1-triv2}
&D_B(j,k)=D_C(j,k) & 
&\text{for $j \in \ang{-\infty:0}$, $k \in \ang{-\infty:+\infty}$}
\end{align}
Equations \eqref{lm-m-mmult-inf1-triv1}--\eqref{lm-m-mmult-inf1-triv2}
cover all but $n$ nonzeros in each of $D_A$, $D_B$, $D_C$.
These remaining nonzeros have 
$i \in \ang{0:+\infty}$, $j \in \ang{0:n}$, $k \in \ang{-\infty:n}$.
Given the $n$ remaining nonzeros in each of $D_A$, $D_B$,
the $n$ remaining nonzeros in $D_C$
can be computed in time $O\bigpa{n^{1.5}}$ and memory $O(n)$.
\end{mylemma}
\begin{figure}[tb]
\centering
\includegraphics{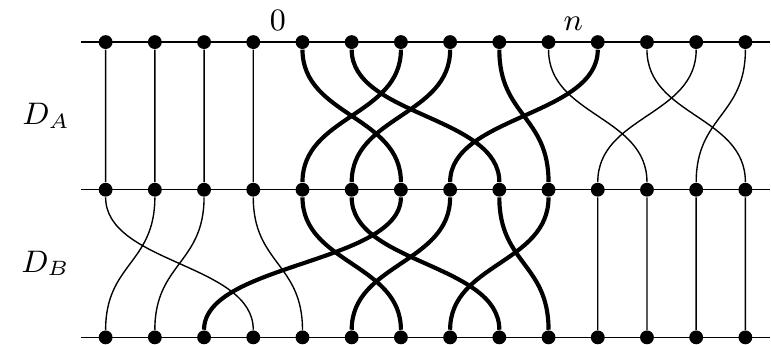}
\caption{\label{f-m-mmult-inf1} 
An illustration of \lmref{lm-m-mmult-inf1}}
\end{figure}
The above lemma is illustrated by \figref{f-m-mmult-inf1}.
Three horizontal lines represent respectively 
the index ranges of $i$, $j$, $k$.
The nonzeros in $D_A$ (respectively, $D_B$) are shown by 
top-to-middle (respectively, middle-to-bottom) seaweeds;
thin seaweeds correspond to the nonzeros covered by 
\eqref{lm-m-mmult-inf1-triv1}--\eqref{lm-m-mmult-inf1-triv2},
and thick seaweeds to the remaining nonzeros.
By \lmref{lm-m-mmult-inf1}, the nonzeros in $D_C$ covered by
\eqref{lm-m-mmult-inf1-triv1}--\eqref{lm-m-mmult-inf1-triv2}
are represented by thin top-to-bottom seaweeds.
The remaining nonzeros in $D_C$ are not represented explicitly,
but can be obtained 
from the thick top-to-middle and middle-to bottom seaweeds
by \lmref{lm-m-mmult}.

\subsection{Partial highest-score matrix multiplication}
\label{s-llcs-partial}

In certain contexts, e.g.\ when $m \gg n$,
we may not be able to solve the all semi-local LCS problem,
or even to store its implicit highest-score matrix.
In such cases, we may wish to settle 
for the following asymmetric version of the problem.
\begin{mydefinition}
\label{def-semilocal-partial}
The \emph{partial semi-local LCS problem}
consists in computing the LCS scores
on substrings of $a$ and $b$ as follows:
\begin{itemize}
\item the \emph{all string-substring LCS problem}:
$a$ against every substring of $b$;
\item the \emph{all prefix-suffix LCS problem}:
every prefix of $a$ against every suffix of $b$;
\item the \emph{all suffix-prefix LCS problem}:
every suffix of $a$ against every prefix of $b$.
\end{itemize}
In contrast with the all semi-local LCS problem,
the comparison of substrings of $a$ against $b$
is not required.
\end{mydefinition}

Let $A$ be the highest-score matrix for the all semi-local LCS problem.
Given an implicit representation of $A$,
the corresponding \emph{partial implicit representation}
consists of all nonzeros $A(i,j)$,
where either $i \in \ang{0:n}$, or $j \in \ang{0:n}$ 
(equivalently, $(i,j) \in 
\ang{0:n} \times \ang{0:+\infty} \cup \ang{-\infty:n} \times \ang{0:n}$).
All such nonzeros are core;
their number is at least $n$ and at most $2n$
(note that the size of a partial implicit representation
is therefore independent of $m$).
The minimum (respectively, maximum) number of nonzeros 
is attained when all (respectively, none of) these nonzeros 
are contained in the submatrix 
defined by $(i,j) \in \ang{0:n} \times \ang{0:n}$.

\begin{mytheorem}
\label{th-range-tree-partial}
Given the partial implicit representation of a highest-score matrix $A$,
there exists a data structure which
\begin{itemize}
\item has size $O\pa{n \log n}$;
\item can be built in time $O\pa{n \log n}$;
\item allows to query an individual element of $A$,
      corresponding to an output of the partial semi-local LCS problem, 
      in time $O\pa{\log^2 n}$.
\end{itemize}
\end{mytheorem}
\begin{myproof}
Similarly to the proof of \thref{th-range-tree},
the structure in question is a 2D range tree
built on the set of nonzeros in the partial implicit representation of $A$.
\myqed
\end{myproof}

The following lemma gives an equivalent of highest-score matrix multiplication
for partially represented matrices.

\begin{mylemma}
\label{lm-score-mmult-partial}
Consider the concatenation of alignment dags 
as described in \ssecref{s-score-mmult},
with highest-score matrices $A$, $B$, $C$.
Given the partial implicit representations of $A$, $B$,
the partial implicit representation of $C$
can be computed in time $O\bigpa{n^{1.5}}$ and memory $O(n)$.
\end{mylemma}
\begin{myproof}
Let $D'_A(i,j)=D_A(i-M,j)$, $D'_B(j,k)=D_B(j,k+m)$,
$D'_C(i,k)=D_B(i-M,k+m)$ for all $i,j,k$,
and define $d'_A$, $d'_B$, $d'_C$ accordingly.
It is easy to check that $d'_A \odot d'_B = d'_C$,
iff $d_A \odot d_B = d_C$.
Matrices $D'_A$, $D'_B$, $D'_C$
satisfy the conditions of \lmref{lm-m-mmult-inf1},
therefore all but $n$ of the core nonzeros
in the required partial implicit representation
can be obtained by 
\eqref{lm-m-mmult-inf1-triv1}--\eqref{lm-m-mmult-inf1-triv2}
in time and memory $O(n)$,
and the remaining $n$ core nonzeros
in time $O(n^{1.5})$ and memory $O(n)$.
\myqed
\end{myproof}

\section{The algorithms}
\label{c-algorithms}

\subsection{Global subsequence recognition and LCS}

We now return to the problem of subsequence recognition 
introduced in \secref{c-slp}.
A simple efficient algorithm for global subsequence recognition
in an SLP-compressed string is not difficult to obtain,
and has been known in folklore%
\footnote{The author is grateful to Y.~Lifshits for pointing this out.}.
For convenience, we generalise the problem's output:
instead of a Boolean value, the algorithm will return an integer.

\begin{myalgorithm}[Global subsequence recognition]
\label{alg-slp-global} \setlabelitbf
\nobreakitem[Input:]
string $T$ of length $m$, represented by an SLP of length $\bar m$;
string $P$ of length $n$, represented explicitly.
\item[Output:]
an integer $k$, giving the length of the longest prefix of $P$
that is a subsequence of $T$.
String $T$ contains $P$ as a subsequence, iff $k=n$.
\item[Description.]
The computation is performed recursively as follows.

Let $T=T' T''$ be the SLP statement defining string $T$.
Let $k'$ be the length of the longest prefix of $P$
that is a subsequence of $T'$.
Let $k''$ be the length of the longest prefix of $P \ldrop k'$ 
that is a subsequence of $T''$.
Both $k'$ and $k''$ can be found recursively.
We have $k=k'+k''$.

The base of the recursion is $\bar m=m=1$.
In this case, the value $k \in \{0,1\}$ 
is determined by a single character comparison.
\item[Cost analysis.]
The running time of the algorithm is $O(\bar mk)$.
The proof is by induction.
The running time of the recursive calls is respectively
$O(\bar mk')$ and $O(\bar mk'')$.
The overall running time of the algorithm
is $O(\bar mk')+O(\bar mk'')+O(1)=O(\bar mk)$.
In the worst case, this is $O(\bar mn)$.
\end{myalgorithm}

We now address the more general partial semi-local LCS problem.
Our approach is based on the technique
introduced in \ssecref{s-llcs-partial}.

\begin{myalgorithm}[Partial semi-local LCS]
\label{alg-slp-lcs} \setlabelitbf
\nobreakitem[Input:]
string $T$ of length $m$, represented by an SLP of length $\bar m$;
string $P$ of length $n$, represented explicitly.
\item[Output:]
the partial implicit highest-score matrix on strings $T$, $P$
\item[Description.]
The computation is performed recursively as follows.

Let $T=T' T''$ be the SLP statement defining string $T$.
Given the partial implicit highest-score matrices 
for each of $T'$ and $T''$ against $P$,
the partial implicit highest-score matrix of $T$ against $P$
can be computed by \lmref{lm-score-mmult-partial}.

The base of the recursion is $\bar m=m=1$.
In this case, the matrix coincides 
with the full implicit highest-score matrix,
and can be computed by a simple scan of string $P$.

\setlabelitbf
\item[Cost analysis.]
By \lmref{lm-score-mmult-partial},
each implicit matrix multiplication 
runs in time $O(n^{1.5})$ and memory $O(n)$.
There are $\bar m$ recursive steps in total,
therefore all the matrix multiplications combined
run in time $O(\bar mn^{1.5})$ and memory $O(n)$.
\end{myalgorithm}

Note that the above algorithm, as a special case, 
provides an efficient solution for the LCS problem:
the LCS score for $T$ against $P$
can easily be queried from the algorithm's output 
by \thref{th-range-tree}.

The running time of \algref{alg-slp-lcs} should be contrasted 
with standard uncompressed LCS algorithms,
running in time $O\bigpa{\frac{mn}{\log(m+n)}}$
\cite{Masek_Paterson:80,Crochemore+:03_SIAM},
and with the NP-hardness of the LCS problem on two compressed strings
\cite{Lifshits_Lohrey:06}.

\subsection{Local subsequence recognition}

We now show how the partial semi-local LCS algorithm of the previous section
can be used to provide local subsequence recognition.

\begin{myalgorithm}[Minimal-window subsequence recognition]
\label{alg-slp-local-min} \setlabelitbf
\nobreakitem[Input:]
string $T$ of length $m$, represented by an SLP of length $\bar m$;
string $P$ of length $n$, represented explicitly.
\item[Output:]
the number of windows in $T$ containing $P$ minimally as a subsequence.
\item[Description.]
\setlabelit
The algorithm runs in two phases.

\setlabelit
\item[First phase.]
Using \algref{alg-slp-lcs}, we compute 
the partial implicit highest-score matrix for every SLP symbol against $P$.
For each of these matrices, we then build
the data structure of \thref{th-range-tree-partial}.

\item[Second phase.]
For brevity, we will call a window containing $P$ minimally as a subsequence
a \emph{$P$-episode window}.
The number of $P$-episode windows in $T$
is computed recursively as follows.

Let $T=T' T''$ be the SLP statement defining string $T$.
Let $m'$, $m''$ be the (uncompressed) lengths of strings $T'$, $T''$.
Let $r'$ (respectively, $r''$) be the number 
of $P$-episode windows in $T'$ (respectively, $T''$),
computed by recursion.

We now need to consider the $n-1$ possible prefix-suffix decompositions 
$P=(P \ltake n')(P \rtake n'')$, for all $n',n''>0$, such that $n'+n''=n$.
Let $l'$ (respectively, $l''$) be the length
of the shortest suffix of $T'$ (respectively, prefix of $T''$)
containing $P \ltake n'$ (respectively, $P \rtake n''$) as a subsequence.
The value of $l'$ (respectively, $l''$) can be found,
or its non-existence established, by binary search 
on the first (respectively, second) index component of nonzeros 
in the partial implicit highest-score matrix 
of $T'$ (respectively, $T''$) against $P$.
In every step of the binary search, we make 
a suffix-prefix (respectively, prefix-suffix) LCS score query 
by \thref{th-range-tree-partial}.
We call the interval $\bra{m'-l':m'+l''}$ a \emph{candidate window}.

It is easy to see that if a window in $T$ is $P$-episode,
then it is either contained within one of $T'$, $T''$, or is a candidate window.
Conversely, a candidate window $\bra{i,j}$ is $P$-episode, 
unless there is a smaller candidate window $\bra{i_1,j_1}$, 
where either $i = i_1 < j_1 < j$, or $i < i_1 < j_1 = j$.
Given the set of all candidate windows 
sorted separately by the lower endpoints and the higher endpoints,
this test can be performed in overall time $O(n)$.
Let $s$ be the resulting number of distinct $P$-episode candidate windows.
The overall number of $P$-episode windows in $T$ is equal to $r'+r''+s$.

The base of the recursion is $m<n$.
In this case, no windows of length $n$ or more exist in $T$, 
so none can be $P$-episode.

\setlabelitbf
\nobreakitem[Cost analysis.]
\setlabelit
\nobreakitem[First phase.]
As in \algref{alg-slp-lcs}, 
the main data structure can be built in time $O(\bar mn^{1.5})$.
The additional data structure of \thref{th-range-tree}
can be built in time $\bar m \cdot O(n \log n) = O(\bar mn \log n)$.

\item[Second phase.]
For each of $n-1$ decompositions $n'+n''=n$,
the binary search performs at most $\log n$
suffix-prefix and prefix-suffix LCS queries, each taking time $O(\log^2 n)$.
Therefore, each recursive step 
runs in time $2n \cdot \log n \cdot O(\log^2 n) = O(n \log^3 n)$.
There are $\bar m$ recursive steps in total,
therefore the whole recursion runs in time $O(\bar mn \log^3 n)$.
It is possible to speed up this phase
by reusing data between different instances of binary search and LCS query;
however, this is not necessary 
for the overall efficiency of the algorithm.

The overall computation cost is dominated 
by the cost of building the main data structure in the first phase,
equal to $O(\bar mn^{1.5})$.
\end{myalgorithm}

\begin{myalgorithm}[Fixed-window subsequence recognition]
\label{alg-slp-local-fix} \setlabelitbf
\nobreakitem[Input:]
string $T$ of length $m$, represented by an SLP of length $\bar m$;
string $P$ of length $n$, represented explicitly;
window length $w$.
\item[Output:]
the number of windows of length $w$ in $T$ containing $P$ as a subsequence.
\item[Description.]
\item[First phase.]
As in \algref{alg-slp-local-min}.

\item[Second phase.]
For brevity, we will call a window of length $w$ 
containing $P$ as a subsequence a \emph{$(P,w)$-episode window}.
The number of $(P,w)$-episode windows in $T$
is computed recursively as follows.

Let $T=T' T''$ be the SLP statement defining string $T$.
Let $m'$, $m''$ be the (uncompressed) lengths of strings $T'$, $T''$.
Let $r'$ (respectively, $r''$) be the number 
of $(P,w)$-episode windows in $T'$ (respectively, $T''$),
computed by recursion.

We now need to consider the $w-1$ windows
that span the boundary between $T'$ and $T''$,
corresponding to strings $(T' \rtake w')(T'' \ltake w'')$, 
for all $w',w''>0$, such that $w'+w''=w$.
We call an interval $\bra{m'-w':m'+w''}$ a \emph{candidate window}.
In contrast with the minimal-window problem, 
we can no longer afford to consider every candidate window individually,
and will therefore need to count them in groups of ``equivalent'' windows.

Let $(i,j)$ (respectively, $(j,k)$) be a nonzero 
in the partial highest-score matrix of $T'$ (respectively, $T''$) against $P$.
We will say such a nonzero is \emph{covered} 
by a candidate window $\bra{m'-w':m'+w''}$,
if $i \in \ang{-m':-m'+w'}$ (respectively, $k \in \ang{m''+n-w:m''+n}$).
We will say that two candidate windows are \emph{equivalent},
if they cover the same set of nonzeros both for $T'$ and $T''$.

Since the number of nonzeros for each of $T'$, $T''$ is at most $n$,
the defined equivalence relation has at most $2n$ equivalence classes.
Each equivalence class corresponds 
to a contiguous segment of values $w'$ (and, symmetrically, $w''$),
and is completely described by the two endpoints of this segment.
Given the set of all the nonzeros,
the endpoint description of all the equivalence classes 
can be computed in time $O(n)$.

For each equivalence class of candidate windows, 
either none or all of them are $(P,w)$-episode; in the latter case, 
we will call the whole equivalence class \emph{$(P,w)$-episode}.
We consider each equivalence class in turn, and pick from it 
an arbitrary representative candidate window $\bra{m'-w':m'+w''}$.
Let $l'$ (respectively, $l''$) be the length
of the longest prefix (respectively, suffix) of $P$
contained in $T' \rtake w'$ (respectively, $T'' \ltake w''$) as a subsequence.
The value of $l'$ (respectively, $l''$) can be found by binary search 
on the second (respectively, first) index component of nonzeros 
in the partial implicit highest-score matrix 
of $T'$ (respectively, $T''$) against $P$.
In every step of the binary search, we make 
a suffix-prefix (respectively, prefix-suffix) LCS score query 
by \thref{th-range-tree-partial}.

It is easy to see that the current equivalence class is $(P,w)$-episode,
iff $l'+l'' \geq n$.
Let $s$ be the total size of $(P,w)$-episode equivalence classes.
The overall number of $(P,w)$-episode windows in $T$ is equal to $r'+r''+s$.

The base of the recursion is $m<w$.
In this case, no windows of length $w$ or more exist in $T$, 
so none can be $(P,w)$-episode.

\setlabelitbf
\item[Cost analysis.]
As in \algref{alg-slp-local-min}, the total cost is dominated 
by the cost of the first phase, equal to $O(\bar mn^{1.5})$.
\end{myalgorithm}

The bounded minimal-window subsequence recognition problem
can be solved by a simple modification of \algref{alg-slp-local-min},
discarding all candidate windows of length greater than $w$.
Furthermore, in addition to counting the windows,
\algrefs{alg-slp-local-min} and \ref{alg-slp-local-fix}
can both be easily modified to report all the respective windows 
at the additional cost of $O(\mathit{output})$.

\section{Conclusions}

We have considered several subsequence recognition problems
for an SLP-compressed text against an uncompressed pattern.
First, we mentioned a simple folklore algorithm 
for the global subsequence recognition problem, running in time $O(\bar mn)$.
Relying on the previously developed framework of semi-local string comparison,
we then gave an algorithm for the partial semi-local LCS problem,
running in time $O(\bar mn^{1.5})$;
this includes the LCS problem as a special case.
A natural question is whether the running time of partial semi-local LCS
(or just LCS) can be improved to match global subsequence recognition.

We have also given algorithms for the local subsequence recognition problem
in its minimal-window and fixed-window versions.
Both algorithms run in time $O(\bar mn^{1.5})$,
and can be easily modified to report all the respective windows 
at the additional cost of $O(\mathit{output})$.
Again, a natural question is whether this running time can be further improved.

Another classical generalisation of both the LCS problem
and local subsequence recognition is 
\emph{approximate matching} (see e.g.\ \cite{Navarro:01}).
Here, we look for substrings in the text that are close
to the pattern in terms of the \emph{edit distance},
with possibly different costs 
charged for insertions/deletions and substitutions.
Once again, we can formulate it as a counting problem
(the \emph{$k$-approximate matching problem}):
counting the number of windows in $T$ 
that have edit distance at most $k$ from $P$.
This problem is considered on LZ-compressed strings
(essentially, a special case of SLP-compression) 
in paper \cite{Karkkainen+:03},
which gives an algorithm running in time $O(\bar mnk)$.
It would be interesting to see if this algorithm can be improved
by using the ideas of the current paper.


\bibliographystyle{plain}
\bibliography{auto,books,string,sort,graph}

\end{document}